\documentclass[article]{jss}

\usepackage{orcidlink,thumbpdf,lmodern}

\usepackage{framed}
\usepackage{amsmath}
\usepackage{amssymb}
\usepackage{amsthm}
\usepackage{Sweave}
\usepackage{algorithm}
\usepackage{algpseudocode}

\newcommand{\fct}[1]{\code{#1()}}

\author{Denis Chetverikov\\ UCLA \And Magne Mogstad\\ University of Chicago \And Paweł Morgen\\LMU Munich \And Joseph Romano\\ Stanford University \AND Azeem Shaikh\\ University of Chicago \And Daniel Wilhelm\\LMU Munich }
\Plainauthor{Denis Chetverikov, Magne Mogstad, Paweł Morgen, Joseph Romano, Azeem Shaikh, Daniel Wilhelm}

\title{\pkg{csranks}: An \proglang{R} Package for Estimation and Inference Involving Ranks}
\Plaintitle{csranks: An R Package for Estimation and Inference Involving Ranks}
\Shorttitle{Estimation and Inference Involving Ranks}

\Abstract{
  This article introduces the \proglang{R} package \pkg{csranks} for estimation and inference involving ranks. First, we review methods for the construction of confidence sets for ranks, namely marginal and simultaneous confidence sets as well as confidence sets for the identities of the tau-best. Second, we review methods for estimation and inference in regressions involving ranks. Third, we describe the implementation of these methods in \pkg{csranks} and illustrate their usefulness in two examples: one about the quantification of uncertainty in the PISA ranking of countries and one about the measurement of intergenerational mobility using rank-rank regressions.
}

\Keywords{confidence sets, ranks, ranking, rank-rank regression, \proglang{R}}
\Plainkeywords{confidence sets, ranks, ranking, rank-rank regression, R}

\Address{
  Denis Chetverikov\\
  Department of Economics\\
  University of California at Los Angeles\\
  315 Portola Plaza\\
  Bunche Hall\\
  Los Angeles, CA 90024, USA\\
  E-mail: \email{chetverikov@econ.ucla.edu}\\ \\
  \emph{and}\\ \\
  Magne Mogstad, Azeem Shaikh\\
  Department of Economics\\
  University of Chicago\\
  1126 East 59th Street\\
  Chicago, IL 60637, USA\\
  E-mail: \email{mogstad@uchicago.edu}, \email{amshaikh@uchicago.edu}\\ \\
  \emph{and}\\ \\
  Paweł Morgen, Daniel Wilhelm\\
  Department of Statistics\\
  Ludwig-Maximilians-Universit\"at M\"unchen\\
  Akademiestr. 1\\
  80799 M\"unchen, Germany\\
  E-mail: \email{seriousmorgen@pm.me}, \email{d.wilhelm@lmu.de}\\ \\
  \emph{and}\\ \\
  Joseph P. Romano\\
  Departments of Economics and Statistics\\
  Stanford University\\
  Sequia Hall\\
  390 Jane Stanford Way\\
  Stanford, CA 94305, USA\\
  E-mail: \email{romano@stanford.edu}
}

\begin{document}

\section[Introduction]{Introduction}
\label{sec: intro}

In economics and other social sciences, it is often desirable to rank populations according to some performance measure or to rank observations before running regressions. For instance, it may be desired to rank neighborhoods according to some measure of intergenerational mobility, countries according to some measure of academic achievement, or hospitals according to patients' average waiting times (\cite{Mogstad:2023aa}). A prominent example for regressions involving ranked observations is the study of intergenerational mobility in which the slope coefficient of a rank-rank regression is a popular measure of the persistence in socioeconomic status across generations (\cite{Deutscher:2023oo} and \cite{Mogstad:2023uu}).

In this paper, we introduce the \proglang{R} package \pkg{csranks} and show how it can be used to perform inference on ranks as well as in regressions involving ranks. 

First, we review the statistical methods proposed by \cite{Mogstad:2023aa} and \cite{Mogstad:2021bb} for the construction of confidence sets for ranks. In this context, there are several populations (e.g., neighborhoods, countries, hospitals) that we want to rank according to some estimated performance measure. Since the performance measure is estimated (e.g., because it is computed on a random sample of data from the population), the statistical uncertainty in the performance measure transfers into statistical uncertainty in the ranking of the populations according to these (estimated) performance measures. \cite{Mogstad:2023aa} propose (i) marginal confidence intervals for the rank of a single population, (ii) simultaneous confidence intervals for the ranks of all populations, and (iii) confidence intervals for the $\tau$-best populations. We show how each of these can be computed using functions from the \pkg{csranks} package. We then provide an empirical illustration using data from the PISA study for ranking countries according to their students' scholastic performance.

Second, we review the statistical methods proposed by \cite{Chetverikov:2023aa} for inference on regressions involving ranks. The main specification, which is popular in empirical work in economics, is a rank-rank regression in which both the independent and the dependent variable are first transformed into ranks and then one of the ranked variables is regressed on the the other, possibly including further (non-ranked) covariates. Inference in such regressions is nonstandard because both the independent and dependent variables have been estimated. The OLS estimator of the regression coefficients is asymptotically equivalent to a U-statistic. \cite{Chetverikov:2023aa} show that the estimator is asymptotically normal and derive the asymptotic variance. In addition, this paper provides asymptotic normality results for several related regression specifications that involve ranked variables. We show how the \pkg{csranks} package can be used to compute all of these estimators and how to perform inference on the regression coefficients, e.g., by computing standard errors and confidence intervals.

Many statistical tests and software packages involve ranks because they consider rank-based statistics, e.g. Wilcoxon's statistic, which are not the subject of this paper. The only statistical software related to inference on ranks that we are aware of is the the \proglang{R} package \pkg{ICRanks} by Al Mohamad et al., based on \cite{mohamad2017simultaneous}, \cite{mohamad2017improvement} and \cite{al2022simultaneous}. It implements a number of alternative methods for construction of confidence sets for ranks. However, the package's scope is restricted to simultaneous confidence sets in the case when the performance measures are independent and follow a Gaussian distribution. In contrast, the methods in the \pkg{csranks} package do not require these two assumptions and, in addition, are weakly more powerful as they implement stepwise improvements. For inference in rank-rank regressions, applied researchers typically employ standard covariance estimators for OLS regressions such as the homoskedastic or heteroskedasticity-robust estimators implemented in the \proglang{R} package \pkg{sandwich} (\cite{zeileis2020various,zeileis2004econometric}). As shown by \cite{Chetverikov:2023aa}, these estimators do not lead to valid inference in rank-rank regressions and the \pkg{csranks} package implements the valid inference procedures proposed by \cite{Chetverikov:2023aa}. The \proglang{R} packages \pkg{copula} (\cite{copula1,copula2,copula3,copula4}) and \pkg{rvinecopulib} (\cite{rvinecopulib}) implement methods for estimation and inference on parameters of copulas. In a special case, the slope of the rank-rank regression is equal to Spearman's rank correlation, a feature of the copula of the independent and dependent variables in the regression, but in general the parameters considered in \pkg{csranks} cannot be written as a feature of the copula and thus \pkg{copula} and \pkg{rvinecopulib} cannot be used for inference in rank-rank regressions.

\section{Overview of the Methods and Implementation}\label{sec:methods}

\subsection{Definition of Ranks}

Suppose we want to create a ranking of $p$ populations, e.g. countries, political parties or hospitals, according to some performance measure $\theta_1,\ldots,\theta_p$. The rank of a population can be defined in different ways. First, ranks can be defined such that the population with the largest performance measure is assigned rank 1, the second largest is assigned rank 2 and so on. We will refer to this as a ``decreasing'' ranking as the rank of population $j$ is decreasing in its own performance measure $\theta_j$. Alternatively, the population with the largest performance measure could be assigned the rank $p$, the second largest is assigned the rank $p-1$ and so on (``increasing'' ranking). Second, if there are ties among the performance measures $\theta_1,\ldots,\theta_p$, then one needs to decide which rank to assign to tied populations. For instance, consider three populations with $\theta_1=10$, $\theta_2=\theta_3=20$. In a decreasing ranking, population 1 is assigned rank $3$. The populations 2 and 3 are tied with the largest performance measure. So, both could be assigned rank 1, both could be assigned rank 2, or any value between 1 and 2.

Let $\theta := (\theta_1,\ldots,\theta_p)'$. A general definition of an increasing rank for population $j$ is
\begin{equation}\label{eq: increasing integer rank}
    R_j^\theta := \omega \sum_{k=1}^p 1\{\theta_k\leq \theta_j\} + (1-\omega)\sum_{k=1}^p 1\{\theta_k < \theta_j\} + 1 -\omega
\end{equation}
where $\omega\in[0,1]$ is a parameter that describes how ties are handled. If none of the other populations are tied with population $j$ ($\theta_j\neq\theta_k$ for all $k$), then the rank of $j$ is equal to $1+ \sum_{k=1}^p 1\{\theta_k< \theta_j\}$ and does not depend on $\omega$. A decreasing rank is obtained by multiplying all performance measures by $-1$:
\begin{equation}\label{eq: decreasing integer rank}
    R_j^{\theta} := \omega \sum_{k=1}^p 1\{\theta_k\geq  \theta_j\} + (1-\omega)\sum_{k=1}^p 1\{\theta_k > \theta_j\} + 1 -\omega
\end{equation}

\begin{table}[t]
    
    \centering

    \begin{tabular}{l|cccccccccc}
    \hline

    \hline
    $j$                                    & 1   & 2   & {\bf 3  } & {\bf 4  } & 5   & 6   & {\bf 7  } & {\bf 8  } & {\bf 9  } & {\bf 10 }\\
    $\theta_j$                                  & 3   & 4   & {\bf 7  } & {\bf 7  } & 10  & 11  & {\bf 15 } & {\bf 15  }& {\bf 15 } & {\bf 15 }\\
    \hline
    smallest rank: $R_j^{\theta}$ for $\omega=0$     & 1 & 2 & {\bf 3} & {\bf 3} & 5 & 6 & {\bf 7} & {\bf 7} & {\bf 7} & {\bf 7} \\
    mid-rank: $R_j^{\theta}$ for $\omega=0.5$      & 1 & 2 & {\bf 3.5} & {\bf 3.5} & 5 & 6 & {\bf 8.5} & {\bf 8.5} & {\bf 8.5} & {\bf 8.5} \\
    largest rank: $R_j^{\theta}$ for $\omega=1$  & 1 & 2 & {\bf 4} & {\bf 4} & 5 & 6 & {\bf 10  } & {\bf 10  } & {\bf 10  } & {\bf 10  }\\
    \hline

    \hline
    \end{tabular}
    \caption{Example of increasing rankings that handle ties differently depending on the value $\omega$. Bold columns indicate ties.}
    \label{tab: example ranks}
\end{table}

Sometimes, it is useful to scale the integer ranks defined above back to the $[0,1]$ interval. A common way to do this in practice is to divide the integer ranks by $p$. For instance, the increasing (``fractional'') rank for population $j$ is then
\begin{equation}\label{eq: increasing fractional rank}
    R_j^\theta := \omega \underbrace{\frac{1}{p}\sum_{k=1}^p 1\{\theta_k\leq \theta_j\}}_{=: \hat{F}_{\theta}(\theta_j)} + (1-\omega)\underbrace{\frac{1}{p}\sum_{k=1}^p 1\{\theta_k < \theta_j\}}_{=: \hat{F}_{\theta}^-(\theta_j)} + \frac{1 -\omega}{p}
\end{equation}
which can be expressed as a weighted average of the empirical cdf $\hat{F}_{\theta}(\theta_j)$ and $\hat{F}_{\theta}^-(\theta_j)$. When $\omega=1$, then the rank corresponds to a popular definition of the rank in applied work, namely the rank of $j$ is equal to the empirical cdf evaluated at population $j$'s performance measure.

\subsection{Confidence Sets for Ranks}
\label{sec: methods}

For concreteness, in this section we consider the ranking of $j=1,\ldots,p$ countries according to how well they educate their children (as in our empirical illustration in Section~\ref{sec: pisa}). The true performance measures for the countries are $\theta_1, \ldots, \theta_p$. These are not observed directly. Instead, for each country we observe data of sample size $n$ from which we compute estimators $\hat{\theta}_1,\ldots,\hat{\theta}_p$ of the performance measures $\theta_1, \ldots, \theta_p$. These estimators may be sample averages of children's test scores, for example. However, the methods below are theoretically justified for the general case in which $\hat{\theta}:=(\hat{\theta}_1,\ldots,\hat{\theta}_p)'$ is a consistent estimator of some parameter $\theta=(\theta_1, \ldots, \theta_p)'$ as long as we can construct a consistent estimator $\hat{\Sigma}$ of the $p\times p$ asymptotic covariance matrix $\Sigma$, with $(j,k)$-element denoted by $\hat{\sigma}_{jk}$, such that
\begin{equation}\label{eq: asy normality}
    \hat{\Sigma}^{-1/2} (\hat{\theta}-\theta) \to_d N(0,I)
\end{equation}
as $n\to\infty$. In this section, we focus on decreasing ranks, i.e., \eqref{eq: decreasing integer rank}, with $\omega=0$. The goal is to use data from each country to form confidence sets $R_{n,j}$ that cover the rank of country $j$, i.e., $R_j^{\theta}$, with probability (approximately) no less than a prespecified level (e.g., $0.95$).

\subsubsection{Marginal Confidence Sets}
\label{sec: marginal}

The goal in this subsection is to construct a two-sided confidence set $R_{n,j}$ for the rank of a particular country $j$ that satisfies\footnote{In fact, the construction described in this section satisfies a stronger property, $\liminf_{n \rightarrow \infty} \inf_{P\in\mathbf{P}} P\left\{R_j^{\theta}(P) \in R_{n,j}\right\} \geq 1 - \alpha$, i.e., asymptotic uniform coverage. For details, see \cite{Mogstad:2023aa}.}
\begin{equation} \label{eq:marginalcoverage}
    \lim_{n \rightarrow \infty} P\left\{R_j^{\theta} \in R_{n,j}\right\} \geq 1 - \alpha
\end{equation}
for some pre-specified confidence level $1-\alpha$. This requirement means that the set $R_{n,j}$ covers the true rank of country $j$ with asymptotic probability no less than $1-\alpha$. The construction is based on simultaneous confidence sets for the differences of performance measures as in \cite{Mogstad:2023aa} and \cite{Mogstad:2021bb}.
For concreteness, we explain one particular approach based on the parametric bootstrap which exploits the asymptotic normality in \eqref{eq: asy normality}, but other constructions are possible; see \cite{Mogstad:2023aa}. To this end consider the confidence set
\begin{equation}
    C_{{\rm symm}, n,j,k} :=  \left[\hat{\theta}_j - \hat{\theta}_k \pm \hat{se}_{jk} c_{{\rm symm}, n,j}^{1-\alpha} \right], \label{eq:Cnsymm}
\end{equation}
where $\hat{se}_{jk}^2 := \hat{\sigma}_{jj} + \hat{\sigma}_{kk}-2\hat{\sigma}_{jk} $ is an estimate of the variance of $\hat{\theta}_j-\hat{\theta}_k$ and $c_{{\rm symm}, n,j}^{1-\alpha}$ is the $(1-\alpha)$-quantile of 
$$\max_{k\colon k\neq j} \frac{|\hat{\theta}_j - \hat{\theta}_k - (\theta_j-\theta_k)|}{\hat{se}_{jk}}.$$
This quantile can be simulated using a parametric bootstrap based on \eqref{eq: asy normality} as follows. Generate $m$ draws of normal random vectors $Z:=(Z_1,\ldots,Z_p)' \sim N(0,\hat{\Sigma}))$. The desired quantile $c_{{\rm symm}, n,j}^{1-\alpha}$ can then be approximated by the empirical ($1-\alpha$)-quantile of the $m$ draws of $\max_{k\colon k\neq j} |Z_j-Z_k|/\hat{se}_{jk}$.

Under weak conditions, the confidence sets for the differences simultaneously cover all true differences involving country $j$:
\begin{equation} \label{eq:coverage}
    \lim_{n \rightarrow \infty}  P\{\theta_j-\theta_k \in C_{{\rm symm}, n,j,k}
    \text{ for all }k \text{ with } k\neq j \} \geq 1 - \alpha.
\end{equation}
Collect the countries $k$ whose differences with $j$ have a confidence set $C_{{\rm symm}, n,j,k}$ that lies entirely below zero,
$$N_j^- := \{ k \colon k\neq j\text{ and } C_{{\rm symm}, n,j,k} \subseteq \mathbf R_-\}, $$
and similarly
$$N_j^+ := \{ k \colon k\neq j\text{ and } C_{{\rm symm}, n,j,k} \subseteq \mathbf R_+\}. $$
Thus $N_j^-$ contains all countries $k$ that have a significantly larger performance measure than $j$, while $N_j^+$ contains all the countries $k$ that have a significantly smaller performance measure than $j$. If the true performance measures of countries $k$ in $N_j^-$ ($N_j^+$) were indeed all larger (smaller) than that of country $j$, then the rank of country $j$ could not be better than $|N_j^-|+1$ and not be worse than $p-|N_j^+|$. Thus, the set 
\begin{equation}\label{eq: Rnj}
  R_{n,j} := \{|N_j^-|+1,\ldots, p-|N_j^+|\}
\end{equation}
would contain the true rank of country $j$. Of course, the confidence sets for the differences cover the true differences only with probability approximately no less than $1-\alpha$, so $R_{n,j}$ covers the true rank of country $j$ only with probability approximately no less than $1-\alpha$. In conclusion, for the construction described in this subsection, \eqref{eq:coverage} implies that $R_{n,j}$ is a confidence set for the rank $R_j^{\theta}$ satisfying \eqref{eq:marginalcoverage} as desired.

It is possible to improve the simple construction of $R_{n,j}$ above by inverting hypotheses tests of 
\begin{equation*}\label{eq:family two-sided}
  H_{j,k}\colon \theta_j-\theta_k= 0
\end{equation*}
versus its negation, for all $k$ that are not equal to $j$. After testing this family of hypotheses, one then counts the number of hypotheses that were rejected in favor of $\theta_j<\theta_k$ and in favor of $\theta_j>\theta_k$. The first number plus one is then used as lower endpoint and the second number subtracted from $p$ is then used as upper endpoint for $R_{n,j}$. This confidence set satisfies \eqref{eq:marginalcoverage} provided that the procedure used to test the family of hypotheses controls the mixed directional familywise error rate (mdFWER) at $\alpha$, i.e., 
$$\lim_{n\to\infty} \text{mdFWER} \leq \alpha,$$
where mdFWER is the probability of making any mistake, either a false rejection or an incorrect determination of a sign; see \cite{Mogstad:2023aa} for details.

\subsubsection{Simultaneous Confidence Sets}
\label{sec: simul}

A small modification of the above construction of a marginal confidence set for the rank of a single country delivers two-sided confidence sets $R_{n,j}$ for the ranks of all countries $j=1,\ldots,p$ such that all true ranks are covered simultaneously, i.e.,
\begin{equation} \label{eq:jointcoverage}
  \lim_{n \rightarrow \infty} P\{R_j^{\theta} \in R_{n,j} \text{ for all } j=1,\ldots,p\} \geq 1 - \alpha.
\end{equation}
We start with confidence sets for the differences $C_{{\rm symm}, n,j,k} $ as in \eqref{eq:Cnsymm} except that the critical value $c_{{\rm symm}, n,j}^{1-\alpha}$ is now defined as the $(1-\alpha)$-quantile of 
$$\max_{(j,k)\colon k\neq j} \frac{|\hat{\theta}_j - \hat{\theta}_k - (\theta_j-\theta_k)|}{\hat{se}_{jk}},$$
where the max is taken over all pairs $(j,k)$ such that $j\neq k$, so the critical value is independent of $j$. As above this critical value can be approximated by the ($1-\alpha$)-quantile of the $m$ draws of $\max_{(j,k)\colon k\neq j} |Z_j-Z_k|/\hat{se}_{jk}$. Then, the confidence set for country $j$, $R_{n,j}$, is computed as in \eqref{eq: Rnj} using the definitions of $N_j^-$, $N_j^+$ as above except that the confidence sets for the differences, $C_{{\rm symm}, n,j,k} $, are replaced by the new ones described here.

Stepwise methods can be used to improve this simple construction of simultaneous confidence sets similarly to the stepwise improvements described for the marginal confidence sets.

\subsubsection{Confidence Sets for the tau-best}
\label{sec: tau-best}

In this section, we are interested in constructing confidence sets for the $\tau$-best countries, defined as 
$$R_0^{\tau-\rm{best}} := \{j\in\{1,\ldots,p\} : R_j^{\theta} \leq \tau \}.$$
This is the set of countries that have the $\tau$ largest performance measures, the ``top-$\tau$'' countries. If there are no ties, then this set contains exactly $\tau$ countries. If there are ties, then the set may contain more countries.

We want to learn which countries could be in this set, i.e. among the top-$\tau$. To this end we construct a (random) set $R^{\tau-\rm{best}}_n$ satisfying
\begin{equation} \label{eq: coverage tau-best}
    \lim_{n \rightarrow \infty} P\left\{R_0^{\tau-\rm{best}} \subseteq R^{\tau-\rm{best}}_n\right\} \geq 1 - \alpha~.
\end{equation}
Such a confidence set contains all the countries that cannot be rejected to be among the top-$\tau$. By construction, the confidence set contains at least $\tau$ countries, but typically more. 

Let $R_{n,j}$, $j=1,\ldots,p$, be simultaneous lower confidence bounds on the ranks of all journals, i.e., each $R_{n,j}$ has upper endpoint equal to $p$ and \eqref{eq:jointcoverage} is satisfied. Such one-sided confidence sets for the ranks can be constructed similarly as the two-sided confidence sets described in Section~\ref{sec: simul}, except that the two-sided confidence sets for the differences are replaced by one-sided confidence sets; see Appendix~\ref{app: one-sided} for details. Then,
\begin{equation}\label{eq: tau-best cs}
  R^{\tau-\rm{best}}_n := \left\{j \in \{1,\ldots,p\} : \tau \in R_{n,j} \right\}
\end{equation}
is a confidence set satisfying \eqref{eq: coverage tau-best}. \cite{Mogstad:2023aa} propose a different, more direct approach to constructing confidence sets for the $\tau$-best, which in simulations has been shown to produce shorter confidence sets, but is computationally more demanding. The \pkg{csranks} package currently only implements the simpler construction in \eqref{eq: tau-best cs}, which is referred to as the projection confidence set.

Confidence sets for the $\tau$-worst can be constructed as confidence sets for the $\tau$-best among $-\theta(P_1),\ldots,-\theta(P_p)$.

\subsubsection{Finite Sample Inference for Multinomial Data}
\label{sec: multinomial}

Consider the special case in which the data come from a poll in which a random sample of respondents are asked to choose one of $p$ political parties. Suppose we want to rank the parties according to $\theta_1,\ldots,\theta_p$, the shares of the total population that support them. Let $X_j$ denote the number of times party $j$ has been chosen by respondents in the poll. Then, $X:=(X_1,\ldots,X_p)'$ is distributed according to the multinomial distribution with parameters $n$, the number of respondents, and $\theta:=(\theta_1,\ldots,\theta_p)'$, the vector of multinomial probabilities.

As above the goal is to form marginal and simultaneous confidence sets for the rank of each party. \cite{Mogstad:2021bb} propose a construction that exploits the multinomial structure of the setup and show that the resulting confidence set $R_{n,j}$ covers the true rank of party $j$ {\it in finite samples}:
\begin{equation} \label{eq: finite sample coverage}
  P\{R_j^{\theta} \in R_{n,j}\} \geq 1 - \alpha
\end{equation}
for any sample size $n$. This is in stark contrast to the asymptotic coverage property of the more general construction in \eqref{eq:marginalcoverage}.

For a given country $j$, the construction of the confidence set is based on hypotheses tests of the family $$H_{k,l}\colon \theta_k\leq \theta_l$$
for all pairs $(k,l)$ such that $k\neq l$ and one of the two is equal to $j$. Then, let
$$R_{n,j} := \left\{|\text{Rej}_{j}^-| + 1, \ldots, p-|\text{Rej}_{j}^+|\right\},$$
where
\begin{equation}\label{eq: rej minus}
  \text{Rej}_{j}^{-} := \{k\neq j\colon \text{ reject } H_{k,{j}} \text{ and claim } \theta_{j}<\theta_k\}
\end{equation}
indicate the set of hypotheses that are rejected in favor of $\theta_{j}<\theta_k$ and 
\begin{equation}\label{eq: rej plus}
  \text{Rej}_{j}^{+} := \{k\neq j \colon \text{ reject } H_{j,k} \text{ and claim } \theta_{j}>\theta_k\}
\end{equation}
the set of hypotheses that are rejected in favor of $\theta_{j}>\theta_k$. If one has a test of this family of hypotheses that controls the familywise error rate (FWER) at $\alpha$, then the resulting confidence set $R_{n,j}$ satisfies $P\{R_j^{\theta} \in R_{n,j}\} \geq 1-FWER \geq 1-\alpha$ as desired. So, the remaining task is to provide tests of the individual hypotheses $H_{k,l}$ such that, overall, the FWER is controlled. 

This is achieved in two steps. First, consider testing a single hypothesis for a given pair $k$ and $l$. Let $S_{k,l}:=X_k+X_l$. One can show that the conditional distribution of $X_k$ given $S_{k,l}=s$ is binomial based on $s$ trials and success probability $\theta_k/(\theta_k+\theta_l)$. This is intuitive because, given that parties $j$ and $k$ together have been chosen by $S_{k,l}=s$ respondents, the distribution of $X_k$ is a binomial experiment about how often out of these $s$ trials $k$ was chosen. Therefore, testing a single hypothesis is equivalent to testing an inequality for a binomial probability, for which a test with finite sample validity can easily be constructed (\cite{Lehmann:2005p3350}). \cite{Mogstad:2021bb} propose the p-value
\begin{equation}\label{eq: p-val}
  \hat{p}_{k,l} :=  \frac{1}{2^{S_{k,l}}}\sum_{i=X_k}^{S_{k,l}} {S_{k,l} \choose i}
\end{equation}
for testing the individual hypothesis $H_{k,l}$. In the second step, the individual p-values are then combined using the Holm procedure so as to control the FWER for the family of hypotheses $H_{k,l}$ with $(k,l)$ such that $k\neq l$ and one of the two is equal to $j$.

Confidence sets that simultaneously cover the true ranks for all parties are constructed in a similar fashion except that now one needs to test the family of hypotheses $H_{k,l}$ for all pairs $(k,l)$ such that $k\neq l$.

\subsection{Regressions Involving Ranks}
\label{sec: inf for ranks}

Slope coefficients in rank-rank regressions are popular measures of intergenerational mobility, for instance in regressions of a child's income rank on their parent's income rank. In this section, we review recent results by \cite{Chetverikov:2023aa} providing the asymptotic theory for coefficients in regressions involving ranks and the inference methods they propose.

\subsubsection{Rank-Rank Regressions}

In this section, we are interested in regression models of the form
\begin{equation}\label{eq: model with controls}
R_Y(Y) = \rho R_X(X) + W'\beta + \varepsilon,\qquad E\left[\varepsilon\begin{pmatrix}
R_{X}(X)\\ W
\end{pmatrix}\right]=0,
\end{equation}
where, for some $\omega\in[0,1]$,
$$R_X(x) := \omega F_X(x) + (1-\omega)F_X^-(x), $$
$F_X$ is the cdf of a random variable $X$, and $F_X^-(x):=P(X<x)$. $R_Y(y)$ is defined analogously based on the same value of $\omega$ as in $R_X(x)$. $W$ is a vector of regressors, $\rho$ and $\beta$ are coefficients of interest.

Suppose we have an i.i.d. sample $\{(Y_i,X_i,W_i)\}_{i=1}^n$ from the distribution of $(Y,X,W)$. In this case, $R_X(x)$ is the probability limit of 
\begin{equation}\label{eq: def rank fn}
    \hat{R}_X(x) := \omega\hat{F}_X(x) + (1-\omega)\hat{F}_X^-(x)+\frac{1-\omega}{n}
\end{equation}
and the increasing fractional rank defined in \eqref{eq: increasing fractional rank} satisfies $R_i^X = \hat{R}_X(X_i) $. Analogously define $\hat{R}_Y(y)$ so that the increasing fractional rank for $Y$ satisfies $R_i^Y = \hat{R}_Y(Y_i) $. We can then estimate the coefficients $\rho$ and $\beta$ from an OLS regression of $R_i^Y$ on $R_i^X$ and $W_i$:
\begin{equation}\label{eq: joint ols estimator}
\begin{pmatrix}
\hat{\rho}\\
\hat{\beta}
\end{pmatrix}=\left(\sum_{i=1}^{n}\begin{pmatrix}
R_i^X\\
W_{i}
\end{pmatrix}\begin{pmatrix}
R_i^X & W_{i}'\end{pmatrix}\right)^{-1}\sum_{i=1}^{n}\begin{pmatrix}
R_i^X\\
W_{i}
\end{pmatrix}R_i^Y,
\end{equation}
\cite{Chetverikov:2023aa} show that, under weak conditions, this estimator is consistent and asymptotically normal,
\begin{equation}\label{eq: an for rank-rank reg}
    \sqrt{n} \begin{pmatrix}
    \hat\rho -\rho\\ \hat\beta-\beta 
\end{pmatrix} \to_d N(0,\Sigma),
\end{equation}
and derive the expression of $\Sigma$. Importantly, the expression of the asymptotic variance $\Sigma$ differs from the probability limits of commonly used variance estimators such as the homoskedastic and Eicker-White variance estimators that software implementations\footnote{For instance, the \fct{lm} command in \proglang{R} or the \code{regress} command in Stata.} of standard OLS regressions report. This is because both the dependent and the independent variable in the rank-rank regression are estimated. The commonly used variance estimators ignoring this additional estimation error thus lead to invalid standard errors and confidence sets. \cite{Chetverikov:2023aa} show that, in fact, these standard errors may be too large or too small depending on the shape of the copula of $Y$ and $X$. Therefore, the invalid standard errors may lead to conservative or misleading inference.

In the special case, in which $X_i$ and $Y_i$ are both drawn from a continuous distribution and $W_i$ includes only a constant, the OLS estimator is equal to Spearman's rank correlation. Otherwise, it is not.

The \fct{lmranks} function implements the OLS estimator \eqref{eq: joint ols estimator} and standard errors, p-values and t-values based on a consistent estimator of the correct asymptotic variance $\Sigma$. The estimate of the asymptotic variance $\Sigma$ can be calculated with the \fct{vcov} method applied to an \fct{lmranks} object. It is used internally for other methods for
\fct{lmranks} objects, such as \fct{summary} or \fct{confint}.

Considerable care was taken for the implementation of this method to be computationally efficient and scalable. The achieved complexity of the implemented algorithm is linearithmic in terms of the number of observations. Appendix \ref{app:technical} contains technical details of the implementations.

\subsubsection{Other Regressions Involving Ranks}

In this section, we briefly describe some variants of the rank-rank regression which are used in empirical work in economics and also implemented in the \fct{lmranks} function.

\paragraph{Rank-rank regressions with clusters.} We consider a population (e.g., the U.S.) that is divided into $n_G$ subpopulations or ``clusters'' (e.g., commuting zones). We are interested in running rank-rank regressions separately within each cluster. The ranks, however, are computed from the distribution of the entire population (e.g., the U.S.). Such regressions are common in studies of intergenerational mobility (e.g., \cite{Chetty:2018iu}), for instance.

Specifically, we consider the model
\begin{equation}\label{eq: model with clusters}
    R_Y(Y) = \sum_{g=1}^{n_G} 1\{G=g\}\left(\rho_g R_X(X) + W'\beta_g\right) + \varepsilon,\qquad E\left[\left.\varepsilon\begin{pmatrix} R_{X}(X)\\ W\end{pmatrix} \right| G\right]=0\; \text{a.s.},
\end{equation}
where $G$ is an observed random variable taking values in $\{1,\ldots,n_G\}$ to indicate the cluster to which an individual belongs. $(G,X,W,Y)$ have distribution $F$ and we continue to denote marginal distributions of $X$ and $Y$ by $F_X$ and $F_Y$. $F_X^-$, $F_Y^-$, $R_X(x)$, and $R_Y(y)$ are also as previously defined, so that $R_X(X)$, for instance, is the rank of $X$ in the entire population, not the rank within a cluster. So, in the model \eqref{eq: model with clusters}, the coefficients $\rho_g$ and $\beta_g$ are cluster-specific, but the ranks $R_Y(Y)$ and $R_X(X)$ are not. In consequence, $\rho_g$ cannot be interpreted as the rank correlation within the cluster $g$.

Let $\{(Y_i,X_i,W_i,G_i)\}_{i=1}^n$ be a random sample from the distribution of $(Y,X,W,G)$. The coefficients $\rho_g$ and $\beta_g$ for cluster $g$ can be consistently estimated by first constructing the ranks $R_i^X$ and $R_i^Y$ and then running an OLS regression of $R_i^Y$ on $R_i^X$ and $W_i$ using only observations from cluster $g$ (i.e., for which $G_i=g$). Denote by $\hat\rho:=(\hat\rho_1,\ldots,\hat\rho_{n_G})'$ and $\hat\beta:=(\hat\beta_1',\ldots,\hat\beta_{n_G}')'$ the vectors of all cluster-specific OLS estimators of $\rho:=(\rho_1,\ldots,\rho_{n_G})'$ and $\beta:=(\beta_1',\ldots,\beta_{n_G}')'$. Then, \cite{Chetverikov:2023aa} show that
\begin{equation*}\label{eq: AN all coeffs with clusters}
    \sqrt{n} \begin{pmatrix} \hat\rho -\rho\\ \hat\beta-\beta \end{pmatrix} \to_d N(0,\Sigma)
\end{equation*}
and derive the expression of the asymptotic variance $\Sigma$.

\paragraph{Regression of a general outcome on a rank.} In this case, we consider a regression model with a general, non-ranked dependent variable $Y$ and a ranked independent variable:
\begin{equation*}\label{eq: model of outcome on rank}
    Y = \rho R_X(X) + W'\beta + \varepsilon,\qquad E\left[\varepsilon\begin{pmatrix}
    R_{X}(X)\\ W
    \end{pmatrix}\right]=0.
\end{equation*}
Let $\{(Y_i,X_i,W_i)\}_{i=1}^n$ be an i.i.d. sample from the distribution of $(Y,X,W)$. \cite{Chetverikov:2023aa} show that the OLS estimator of a regression of $Y_i$ on $R_i^X$ and $W_i$ is asymptotically normal as in \eqref{eq: asy normality} and derive the expression of the asymptotic variance $\Sigma$. 

\paragraph{Regression of a rank on a general regressor.} In this case, we consider a regression model with a ranked dependent variable and a general, non-ranked independent variable:
\begin{equation*}\label{eq: model of rank on regressor}
    R_Y(Y) = W'\beta + \varepsilon,\qquad E\left[\varepsilon W \right]=0.
\end{equation*}
Let $\{(Y_i,X_i,W_i)\}_{i=1}^n$ be an i.i.d. sample from the distribution of $(Y,X,W)$. \cite{Chetverikov:2023aa} show that the OLS estimator of a regression of $R^Y_i$ on $W_i$ is asymptotically normal,
$$\sqrt{n}(\hat\beta-\beta)\to_d N(0,\Sigma) $$
and derive the expression of the asymptotic variance $\Sigma$.

\section{The package csranks}

The package \pkg{csranks} comprises 14 functions. These functions implement methods for construction of confidence sets for ranks and inference in rank-rank regressions, described in Section \ref{sec:methods}.

\subsection{Inference on ranks}

The central functions for constructing confidence sets for ranks (as described in Sections \ref{sec: marginal}-\ref{sec: simul}) and confidence sets for the $\tau$-best/worst (as described in Section \ref{sec: tau-best}) are \fct{csranks}, \fct{cstaubest} and \fct{cstauworst}. They all require
a vector of estimates \code{x} and an estimate of their covariance matrix \code{Sigma} (if the estimates are independent, the user should pass a diagonal matrix). The nominal coverage of confidence set can be specified with the argument \code{coverage} and the number of bootstrap samples is set with the argument \code{R}. \fct{csranks} also accepts a boolean \code{simul} argument, which specifies whether the returned set should be marginal (\code{FALSE}) or simultaneous (\code{TRUE}) confidence set.

The \fct{csranks\_multinom} function is similar to \fct{csranks}, but designed for the special case in which the data is multinomial (as described in Section~\ref{sec: multinomial}). In the multinomial case, the computation of the covariance matrix of the estimates \code{x} requires only knowledge of \code{x} and thus the function does not require the argument \code{Sigma}. The arguments \code{coverage} and \code{simul} play an identical role as in \fct{csranks}. The only new argument, \code{multcorr}, specifies the method used for  correction of the p-values for multiple testing (\code{Bonferroni} or \code{Holm}).

For both \fct{csranks} and \fct{csranks\_multinom} an S3 \code{plot} method is implemented. Additionally, there are utility functions \fct{irank} and \fct{frank} used to compute integer and fractional ranks, and \fct{irank\_against} and \fct{frank\_against} to compute ranks of one vector based on values in another reference vector.

\subsection{Inference for regressions involving ranks}

The main function for inference in regressions involving ranks (as described in Section \ref{sec: inf for ranks}) is \fct{lmranks}. It is designed to be as similar to the well-known \proglang{R} function \fct{lm} (for linear regressions) as possible. As for \fct{lm}, the most important argument of \fct{lmranks} is the \code{formula} argument, which specifies the model using the \proglang{R} formula syntax (\cite{RManual}). A typical model has the form \code{response ~ terms} where \code{response} is the (numeric) response vector and \code{terms} is a series of terms which specifies a linear predictor for the response. Terms can be added with \code{+}, removed with \code{-}, and a colon \code{:} is used to specify an interaction. 

A new functionality in \fct{lmranks} is that the user can specify variables in the formula to be ranked (i.e. their values be replaced by their ranks) before running the regression. This is achieved by wrapping the response or one of the terms with \code{r()}. A typical rank-rank regression model with ranked response \code{Y}, ranked regressor \code{X}, an intercept and non-ranked regressors \code{W1} and \code{W2} is specified with a formula \code{r(Y) ~ r(X) + W1 + W2}.

The \code{weights} argument is not supported due to lack of theory on weighted rank-rank regression and the \code{subset} and \code{na.action} arguments are not supported due to the order in which the \proglang{R} function \code{model.frame} processes the arguments. It first evaluates the \code{formula}, and then applies the \code{subset} and \code{na.action} arguments  (\cite{RManual}). Since those arguments remove observations from the dataset, it affects the distribution of ranks in the resulting dataset and thus is not permitted. The user is therefore required to subset the data and handle the \code{NA} values on their own, before passing data to \fct{lmranks}.

Many functions defined for \fct{lm} also work correctly with \fct{lmranks}. These include \fct{coef}, \fct{model.frame}, \fct{model.matrix}, \fct{resid}, \fct{predict}, \fct{update} and others. On the other hand, some would return incorrect results if they treated \fct{lmranks} output in the same way as \fct{lm}'s and have been disabled. These functions in most cases require the number of degrees of freedom of the model, and for rank-rank regressions it is not yet clear how to calculate them. The central contribution of this package are \fct{vcov}, \fct{summary} and \fct{confint} implementations using the correct asymptotic theory for regressions involving ranks.

Sometimes, the dataset is divided into clusters and one is interested in running rank-rank regressions separately within each cluster, where the ranks are not computed within each cluster, but using all observations pooled across all cluster. This is the model in \eqref{eq: model with clusters}. This regression model can be written as a rank-rank regression in which the regressors are multiplied by cluster indicators. For $q$ regressors and $n_G$ clusters we get $q n_G$ columns -- one for each regressor-cluster pair. This expansion is conveniently achieved by using functions already implemented in base \proglang{R}. In the \proglang{R} formula, an interaction operator has to be used, and the rest is done in \code{model.matrix}. In \fct{lmranks}, a typical rank-rank regression with clusters specified in the variable \code{G}, ranked response \code{Y}, ranked regressor \code{X}, an intercept and non-ranked regressors \code{W1} and \code{W2} is specified with a formula \code{r(Y) ~ (r(X) + W1 + W2):G}.

\section{Empirical Applications}

\subsection{Ranking Countries by Academic Achievement}
\label{sec: pisa}

The following example illustrates how the \pkg{csranks} package can be used to quantify the statistical uncertainty in the PISA ranking of countries. Over the past two decades, the Organization for Economic Co-operation and Development (OECD) have conducted the PISA study. The goal of this study is to evaluate and compare educational systems across countries by measuring 15-year-old school students’ scholastic performance on math, science, and reading. Each country that participates in a given year draws a sample of students to be tested. The OECD then processes the test results so as to produce a score for each country and publishes league tables ranking countries by their scores.

In this example, we use publicly available data from the 2018 PISA study to examine in which countries school students do best and worst at math.

\subsubsection{Setup}

First, we load the required libraries and the dataset \code{pisa}, which is part of the \pkg{csranks} package. It contain the three test scores and accompanying standard errors for each country (``jurisdiction''):

\begin{CodeChunk}
\begin{CodeInput}
> library(csranks)
> library(ggplot2)
> set.seed(100)
> data(pisa)
> head(pisa)

  jurisdiction science_score science_se reading_score reading_se math_score
1    Australia      502.9646   1.795398      502.6317   1.634343   491.3600
2      Austria      489.7804   2.777395      484.3926   2.697472   498.9423
3      Belgium      498.7731   2.229240      492.8644   2.321973   508.0703
4       Canada      517.9977   2.153651      520.0855   1.799716   512.0169
5        Chile      443.5826   2.415280      452.2726   2.643766   417.4066
6     Colombia      413.3230   3.052402      412.2951   3.251344   390.9323
   math_se
1 1.939833
2 2.970999
3 2.262662
4 2.357476
5 2.415888
6 2.989559
\end{CodeInput}
\end{CodeChunk}

The PISA study’s math scores are stored in \code{math\_score} and their standard errors in \code{math\_se}. The following graph shows the raw math scores with 95\% marginal confidence intervals:

\begin{CodeChunk}
\begin{CodeInput}
> gpl <- ggplot(pisa, aes(x=reorder(jurisdiction,math_score,decreasing=TRUE), 
+                  y=math_score)) + 
+     geom_errorbar(aes(ymin=math_score-2*math_se, ymax=math_score+2*math_se)) +
+     geom_point() + 
+     theme_minimal() +
+     labs(y="math score", x="", title="2018 PISA math score for OECD countries", 
+          subtitle="(with 95\% marginal confidence intervals)") +
+     theme(axis.text.x = element_text(angle = 45, hjust = 1))
>
> ggsave("pisa_math_absolute.pdf", gpl, width=7, height=5)
\end{CodeInput}
\end{CodeChunk}

\begin{figure}[t!]
\centering
\includegraphics{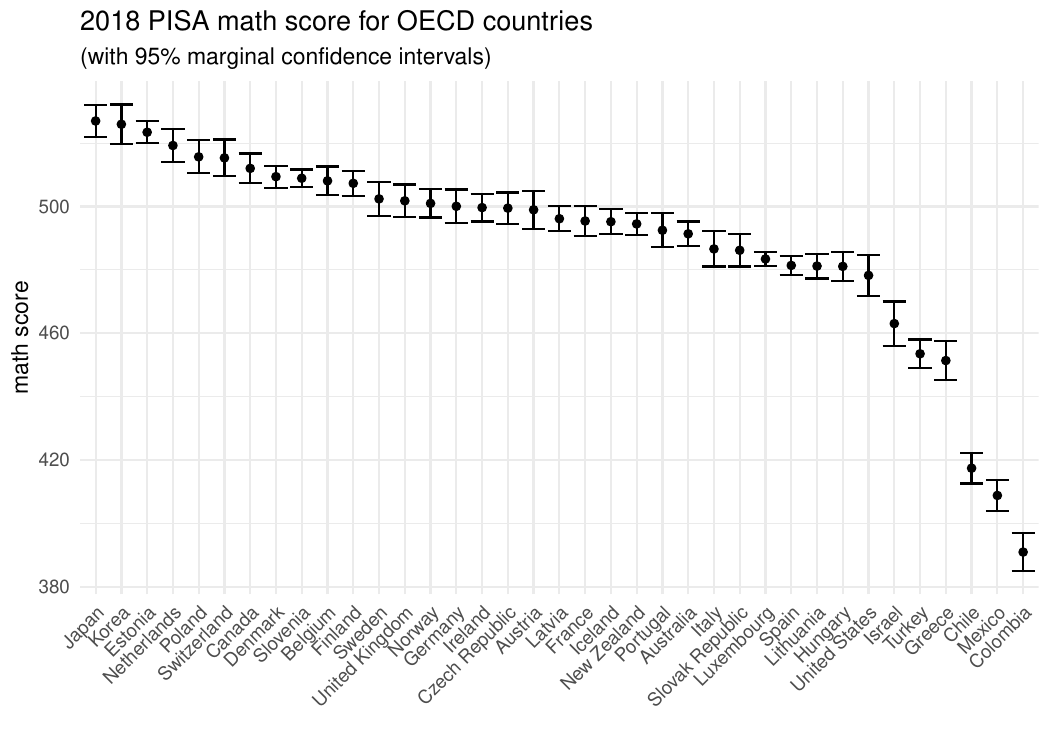}
\caption{Raw math scores from 2018 PISA study.}
\label{fig:pisa_raw}
\end{figure}

Figure~\ref{fig:pisa_raw} shows the resulting graph. The function \fct{irank} can be used to produce integer ranks based on these math scores:

\begin{CodeChunk}
\begin{CodeInput}
> math_rank <- irank(pisa$math_score)
> head(pisa[order(math_rank),c("jurisdiction", "math_score", "math_se")])
\end{CodeInput}
\begin{Soutput}
   jurisdiction math_score  math_se
19        Japan   526.9733 2.471475
20        Korea   525.9330 3.121394
9       Estonia   523.4146 1.743602
25  Netherlands   519.2310 2.632278
28       Poland   515.6479 2.602085
34  Switzerland   515.3147 2.908004

\end{Soutput}
\end{CodeChunk}

Japan is ranked first (i.e., best), Korea is ranked second and so on. Since the math scores are estimates of countries’ true achievements, the ranks assigned to these countries are also estimates, rather than the true ranks. Just like the test scores, the ranks therefore also contain statistical uncertainty. Various functions in the \pkg{csranks} package implement methods for the quantification of this uncertainty, which were described in section \ref{sec:methods}.

\subsubsection{Best populations}

Suppose, that the researcher is interested in finding out which countries could be among the top-5 in terms of their true math score. One can answer this question by constructing the $\tau$-best confidence sets, described in section \ref{sec: tau-best}. In \pkg{csranks}, it is implemented in function \fct{cstaubest}. It requires as an argument an estimate of the covariance matrix of the test scores. In this example, it is assumed the estimates from the different countries are mutually independent, so the covariance matrix is diagonal. The function \fct{cstaubest} can then be used to compute a 95\% confidence set for the top-5:

\begin{CodeChunk}
\begin{CodeInput}
> math_cov_mat <- diag(pisa$math_se^2)
> CS_5best <- cstaubest(pisa$math_score, math_cov_mat, tau = 5, coverage = 0.95)
> pisa[CS_5best, "jurisdiction"]
\end{CodeInput}
\begin{Soutput}
 [1] Belgium     Canada      Denmark     Estonia     Finland     Japan      
 [7] Korea       Netherlands Poland      Slovenia    Sweden      Switzerland
37 Levels: Australia Austria Belgium Canada Chile Colombia ... United States
\end{Soutput}
\end{CodeChunk}

The confidence set contains 12 countries: with probability approximately 0.95, these 12 countries could all be among the top-5 according to their true math score. According to the estimated test scores, the countries Japan, Korea, Estonia, Netherlands, and Poland are the top-5 countries. However, due to the statistical uncertainty in the ranking, there is uncertainty about which countries are truly among the top-5.

\subsubsection{Marginal confidence sets}

Suppose that the researcher is interested in a single country, for example the United Kingdom. She would like to learn where its true ranking may lie. A marginal confidence set, described in section \ref{sec: marginal} and implemented in the function \fct{csranks} with parameter \code{simul=FALSE}, is a way to answer this question:

\begin{CodeChunk}
\begin{CodeInput}
> uk_i <- which(pisa$jurisdiction == "United Kingdom")
> CS_marg <- csranks(pisa$math_score, math_cov_mat, simul=FALSE, 
>                    indices = uk_i, coverage=0.95)
> CS_marg
\end{CodeInput}
\begin{Soutput}
$L
[1] 7

$rank
[1] 13

$U
[1] 23

attr(,"class")
[1] "csranks"
\end{Soutput}
\end{CodeChunk}

\code{CS\_marg\$L} and \code{CSmarg\$U} contain the lower and upper bounds of the confidence set for the rank of the United Kingdom. 

Based on the estimated math scores, the United Kingdom is ranked at 13-th place. However, due to statistical uncertainty in the ranking, its true rank could be anywhere between 7 and 23, with probability approximately 95\%.

\subsubsection{Simultaneous confidence sets}

Finally, suppose the researcher is interested in the entire ranking of countries. Simultaneous confidence sets (described in Section \ref{sec: simul}) for the ranks of all countries quantify the statistical uncertainty in the entire ranking. They are implemented in the function \fct{csranks} with parameter \code{simul=TRUE}:

\begin{CodeChunk}
\begin{CodeInput}
> CS_simul <- csranks(pisa$math_score, math_cov_mat, 
+                     simul=TRUE, coverage=0.95)
> gpl <- plot(CS_simul, popnames=pisa$jurisdiction, 
+      title="Ranking of OECD Countries by 2018 PISA Math Score", 
+      subtitle="(with 95\% simultaneous confidence sets)")
>
> ggsave("pisa_math.pdf", gpl, width=7, height=7)
\end{CodeInput}
\end{CodeChunk}

\begin{figure}[t]
\centering
\includegraphics{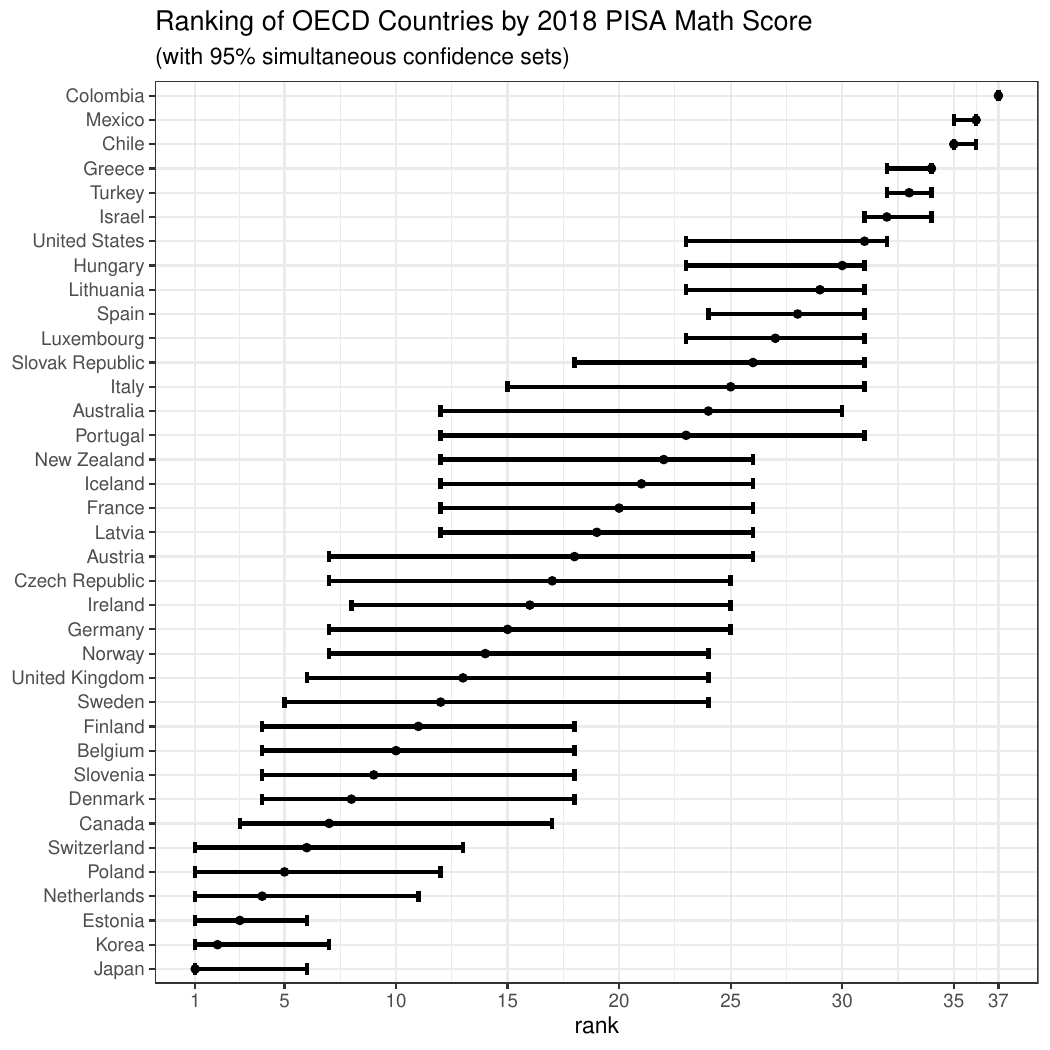}
\caption{Simultaneous confidence sets for ranking according to PISA math test}
\label{fig: sim cs pisa math}
\end{figure}

The resulting graph is shown in Figure~\ref{fig: sim cs pisa math}. The simultaneous confidence sets indicate substantial statistical uncertainty about ranks in the middle of the ranking. For instance, the confidence set for the true rank of Germany has a lower bound of 7 and and upper bound of 24. At the top and the bottom of the ranking, the statistical uncertainty is smaller. For instance, with approximately 95\% probability, the true rank of Colombia is 37. The confidence sets for Mexico and Chile are also very tight and only contain two values.

\subsection{Intergenerational Mobility}

The following example illustrates how the \pkg{csranks} package can be used for estimation and inference in rank-rank regressions. These are commonly used for studying intergenerational mobility.

The dataset used in this example is the \code{parent_child_income} dataset that is part of the \pkg{csranks} package. It is a simulated dataset using a data-generating process calibrated to the National Longitudinal Survey of Youth 1979 from the U.S. Bureau of Labor Statistics. It includes data about parents' (column \code{c\_faminc}) and children's (\code{p\_faminc}) family income, as well as individual characteristics (\code{gender} and \code{race}: \code{"hisp"} (Hispanic), \code{"black"} or \code{"neither"}). 

First, we take a quick look at the dataset:

\begin{CodeChunk}
\begin{CodeInput}
> data(parent_child_income)
> head(parent_child_income)
\end{CodeInput}
\begin{Soutput}
   c_faminc  p_faminc gender    race
1  78771.72 77127.484 female neither
2  79268.33 62723.303 female neither
3  45405.98 65751.340   male neither
4  81951.64 58723.050   male neither
5  88350.33  6381.047   male neither
6 161331.33 40325.466 female neither
> 
\end{Soutput}
\end{CodeChunk}

A popular approach to measuring income mobility is to estimate a rank-rank regression of child's income (\code{c\_faminc}) on a constant and parent's income (\code{p\_faminc}). The \fct{lmranks} function implements this regression. The model can be specified through a formula in which the variables to be ranked are marked by \code{r()}.

\begin{CodeChunk}
\begin{CodeInput}
> lmr_model <- lmranks(r(c_faminc) ~ r(p_faminc), data=parent_child_income)
> summary(lmr_model)
\end{CodeInput}
\begin{Soutput}
The number of residual degrees of freedom is not correct.
Also, z-value, not t-value, since the distribution used for p-value 
calculation is standard normal.

Call:
lmranks(formula = r(c_faminc) ~ r(p_faminc), data = parent_child_income)

Residuals:
     Min       1Q   Median       3Q      Max 
-0.65601 -0.21986 -0.00376  0.22088  0.66495 

Coefficients:
            Estimate Std. Error t value Pr(>|t|)    
(Intercept) 0.312311   0.007161   43.61   <2e-16 ***
r(p_faminc) 0.375538   0.014319   26.23   <2e-16 ***
---
Signif. codes:  0 ‘***’ 0.001 ‘**’ 0.01 ‘*’ 0.05 ‘.’ 0.1 ‘ ’ 1

Residual standard error: NA on 3892 degrees of freedom
\end{Soutput}
\end{CodeChunk}

This regression specification takes each child's income, computes its rank among all children's incomes, then takes each parent's income and computes its rank among all parents' incomes. Then the child's rank is regressed on the parent's rank using OLS. The \fct{summary} method computes standard errors, t-values and p-values according to the asymptotic theory developed in (\cite{Chetverikov:2023aa}). 

One can also run the rank-rank regression with additional covariates, e.g.:

\begin{CodeChunk}
\begin{CodeInput}
> lmr_model_cov <- lmranks(r(c_faminc) ~ r(p_faminc) + gender + race, 
+                       data=parent_child_income)
> summary(lmr_model_cov)
\end{CodeInput}
\begin{Soutput}
The number of residual degrees of freedom is not correct.
Also, z-value, not t-value, since the distribution used for p-value 
calculation is standard normal.

Call:
lmranks(formula = r(c_faminc) ~ r(p_faminc) + gender + race, 
    data = parent_child_income)

Residuals:
     Min       1Q   Median       3Q      Max 
-0.66140 -0.20654 -0.00343  0.21421  0.72917 

Coefficients:
             Estimate Std. Error t value Pr(>|t|)    
(Intercept)  0.299823   0.018155  16.514  < 2e-16 ***
r(p_faminc)  0.323785   0.015123  21.411  < 2e-16 ***
gendermale   0.010862   0.008484   1.280  0.20042    
raceblack   -0.088215   0.020910  -4.219 2.46e-05 ***
raceneither  0.055726   0.018781   2.967  0.00301 ** 
---
Signif. codes:  0 ‘***’ 0.001 ‘**’ 0.01 ‘*’ 0.05 ‘.’ 0.1 ‘ ’ 1

Residual standard error: NA on 3889 degrees of freedom  
\end{Soutput}
\end{CodeChunk}

In some economic applications, it is desired to run rank-rank regressions separately in subgroups of the population, but compute the ranks in the whole population. For instance, we might want to estimate rank-rank regression slopes as measures of intergenerational mobility separately for males and females, but the ranking of children's incomes is formed among all children (rather than form separate rankings for males and females). 

Such regressions can be run using the \fct{lmranks} function with interaction notation:

\begin{CodeChunk}
\begin{CodeInput}
> grouped_lmr_model <- lmranks(r(c_faminc) ~ r(p_faminc):gender,
+                     data=parent_child_income)
> summary(grouped_lmr_model)
\end{CodeInput}
\begin{Soutput}
The number of residual degrees of freedom is not correct.
Also, z-value, not t-value, since the distribution used for p-value 
calculation is standard normal.

Call:
lmranks(formula = r(c_faminc) ~ r(p_faminc):gender, data = parent_child_income)

Residuals:
     Min       1Q   Median       3Q      Max 
-0.65016 -0.21977 -0.00308  0.21750  0.68833 

Coefficients:
                         Estimate Std. Error t value Pr(>|t|)    
genderfemale              0.28796    0.01119   25.74   <2e-16 ***
gendermale                0.33506    0.01107   30.27   <2e-16 ***
r(p_faminc):genderfemale  0.40800    0.02046   19.94   <2e-16 ***
r(p_faminc):gendermale    0.34516    0.02044   16.89   <2e-16 ***
---
Signif. codes:  0 ‘***’ 0.001 ‘**’ 0.01 ‘*’ 0.05 ‘.’ 0.1 ‘ ’ 1

Residual standard error: NA on 3890 degrees of freedom  
\end{Soutput}
\end{CodeChunk}

In this example, we have run a separate OLS regression of children’s ranks on parents’ ranks among the female and male children. However, incomes of children are ranked among all children and incomes of parents are ranked among all parents. The standard errors, t-values and p-values are implemented according to the asymptotic theory developed in (\cite{Chetverikov:2023aa}), where it is shown that the asymptotic distribution of the estimators now need to not only account for the fact that ranks are estimated, but also for the fact that estimators are correlated across gender subgroups because they use the same estimated ranking.

One can also create more granular subgroups by interacting several characteristics such as gender and race:

\begin{CodeChunk}
\begin{CodeInput}
>     parent_child_income$subgroup <- interaction(parent_child_income$gender,
+                                                  parent_child_income$race)
>     gran_grouped_lmr_model <- lmranks(r(c_faminc) ~ r(p_faminc):subgroup,
+                         data=parent_child_income)
>     summary(gran_grouped_lmr_model)
\end{CodeInput}
\begin{Soutput}
The number of residual degrees of freedom is not correct.
Also, z-value, not t-value, since the distribution used for p-value 
calculation is standard normal.

Call:
lmranks(formula = r(c_faminc) ~ r(p_faminc):subgroup, data = parent_child_income)

Residuals:
     Min       1Q   Median       3Q      Max 
-0.65574 -0.20769 -0.00387  0.21297  0.73248 

Coefficients:
                                   Estimate Std. Error t value Pr(>|t|)    
subgroupfemale.hisp                 0.26280    0.03268   8.041 8.94e-16 ***
subgroupmale.hisp                   0.35313    0.04516   7.820 5.26e-15 ***
subgroupfemale.black                0.19741    0.01968  10.031  < 2e-16 ***
subgroupmale.black                  0.26496    0.02702   9.805  < 2e-16 ***
subgroupfemale.neither              0.35096    0.01525  23.006  < 2e-16 ***
subgroupmale.neither                0.36603    0.01322  27.694  < 2e-16 ***
r(p_faminc):subgroupfemale.hisp     0.41733    0.07602   5.490 4.02e-08 ***
r(p_faminc):subgroupmale.hisp       0.21780    0.09697   2.246 0.024694 *  
r(p_faminc):subgroupfemale.black    0.30798    0.05154   5.976 2.29e-09 ***
r(p_faminc):subgroupmale.black      0.25528    0.07234   3.529 0.000417 ***
r(p_faminc):subgroupfemale.neither  0.33908    0.02532  13.390  < 2e-16 ***
r(p_faminc):subgroupmale.neither    0.31818    0.02270  14.018  < 2e-16 ***
---
Signif. codes:  0 ‘***’ 0.001 ‘**’ 0.01 ‘*’ 0.05 ‘.’ 0.1 ‘ ’ 1

Residual standard error: NA on 3882 degrees of freedom
\end{Soutput}
\end{CodeChunk}

Finally, we compare the confidence intervals for rank-rank regression coefficients produced by \fct{lmranks} with those of the naive approach which computes the ranks, then runs a regression of the child's income rank on the parent's income rank using \fct{lm}, and then reports the confidence intervals based on the standard errors from \fct{lm}.

\begin{CodeChunk}
\begin{CodeInput}
>     # compute child's rank
>     c_faminc_rank <- frank(parent_child_income$c_faminc, omega=1, increasing=TRUE)
> 
>     # compute parent's rank
>     p_faminc_rank <- frank(parent_child_income$p_faminc, omega=1, increasing=TRUE)
>     
>     # naive rank-rank regression
>     lm_model <- lm(c_faminc_rank ~ p_faminc_rank)
> 
>     # naive grouped rank-rank regression
>     grouped_lm_model <- lm(c_faminc_rank ~ p_faminc_rank:subgroup + 
+                            subgroup - 1, 
+                            data=parent_child_income)
> 
>     # combine results 
>     theme_set(theme_minimal())
>     ci_data <- data.frame(estimate=coef(lmr_model), 
+                         parameter=c("Intercept", "slope"),
+                         group="Whole sample",
+                         method="csranks", 
+                         lower=confint(lmr_model)[,1], 
+                         upper=confint(lmr_model)[,2])
>     ci_data <- rbind(ci_data, data.frame(
+         estimate = coef(gran_grouped_lmr_model),
+         parameter = rep(c("Intercept", "slope"), each=6),
+         group = rep(c("Hispanic female", "Hispanic male", "Black female", 
+                       "Black male", "Other female", "Other male"), times=2),
+         method="csranks",
+         lower=confint(gran_grouped_lmr_model)[,1],
+         upper=confint(gran_grouped_lmr_model)[,2]
+     ))
>     ci_data <- rbind(ci_data, data.frame(
+         estimate = coef(lm_model),
+         parameter = c("Intercept", "slope"),
+         group = "Whole sample",
+         method="naive",
+         lower=confint(lm_model)[,1],
+         upper=confint(lm_model)[,2]
+     ))
>     ci_data <- rbind(ci_data, data.frame(
+         estimate = coef(grouped_lm_model),
+         parameter = rep(c("Intercept", "slope"), each=6),
+         group = rep(c("Hispanic female", "Hispanic male", "Black female", 
+                       "Black male", "Other female", "Other male"), times=2),
+         method="naive",
+         lower=confint(grouped_lm_model)[,1],
+         upper=confint(grouped_lm_model)[,2]
+     ))
> 
>     # plot confidence sets
>     gpl <- ggplot(ci_data, aes(y=estimate, x=group, ymin=lower, 
+                         ymax=upper,col=method, fill=method)) +
+       geom_point(position=position_dodge2(width = 0.9)) +
+       geom_errorbar(position=position_dodge2(width = 0.9)) +
+       geom_hline(aes(yintercept=estimate), 
+                  data=subset(ci_data, group=="Whole sample"),
+                  linetype="dashed",
+                   col="gray") +
+       coord_flip() +
+       labs(title="95% confidence intervals of intercept and slope
+                   in rank-rank regression") +
+       facet_wrap(~parameter)
> 
>     ggsave("mobility.pdf", gpl)
\end{CodeInput}
\end{CodeChunk}

\begin{figure}[t]
 \centering
 \includegraphics{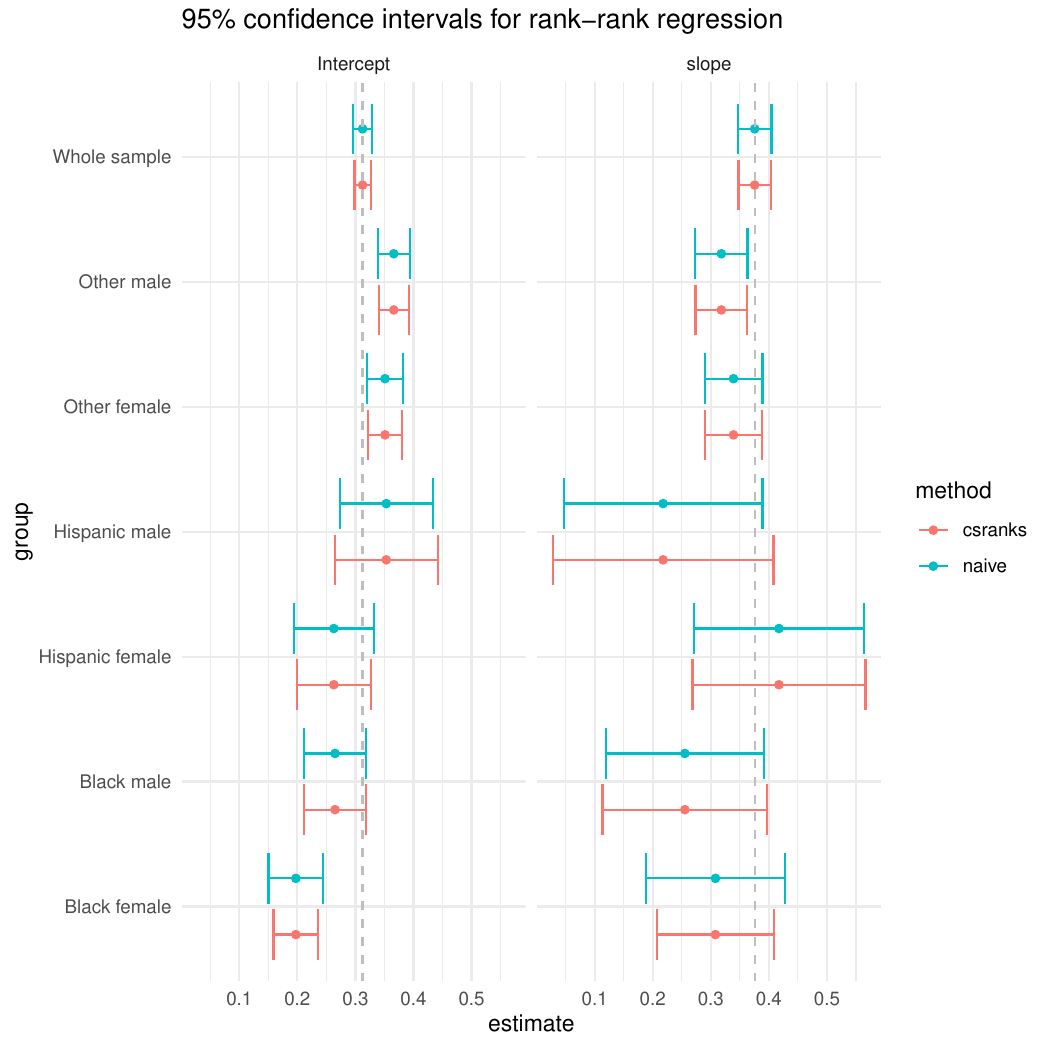}
 \caption{Confidence sets for the intercept and slope parameters in a rank-rank regression on the whole sample and on subgroups. ``csranks'' indicates the valid confidence interval computed through \fct{lmranks} and ``naive'' the confidence intervals from the naive use of \fct{lm} that ignore the estimation error in the ranks.}
 \label{fig:rrreg-cis}
 \end{figure}

Figure~\ref{fig:rrreg-cis} shows the resulting graph comparing the confidence sets obtained from \fct{lmranks} (denoted by ``csranks'') with those of the naive use of \fct{lm} (denoted by ``naive'') that ignores the estimation error in the ranks. ``Whole sample'' refers to the confidence sets for intercept and slope when a rank-rank regression is run on the whole sample. The other rows show the confidence sets for the intercept and slope for each group in the grouped rank-rank regression.

The point estimates are the same for both methods, but the confidence intervals of the naive method are not valid. This is because the usual OLS formulas for standard errors do not take into account the estimation uncertainty in the ranks. This leads to different confidence intervals. The differences in the confidence intervals are not particularly large in this dataset, but \cite{Chetverikov:2023aa} provide further examples in which the differences are considerable.

\section*{Computational details}

The results in this paper were obtained using
\proglang{R}~4.2.1 with the
\pkg{csranks}~1.2.2 and \pkg{ggplot2}~3.4.3 packages. \proglang{R} itself
and all packages used are available from the Comprehensive
\proglang{R} Archive Network (CRAN) at
\url{https://CRAN.R-project.org/}.

\section*{Acknowledgments}

The authors gratefully acknowledge financial support from the European Research Council (Starting Grant No. 852332). Chetverikov, Mogstad, Romano, Shaikh, Wilhelm developed the statistical methods presented in this paper. Morgen and Wilhelm developed the software and wrote the article.

\bibliography{ref}

\begin{thebibliography}{19}
\newcommand{\enquote}[1]{``#1''}
\providecommand{\natexlab}[1]{#1}
\providecommand{\url}[1]{\texttt{#1}}
\providecommand{\urlprefix}{URL }
\expandafter\ifx\csname urlstyle\endcsname\relax
  \providecommand{\doi}[1]{doi:\discretionary{}{}{}#1}\else
  \providecommand{\doi}{doi:\discretionary{}{}{}\begingroup
  \urlstyle{rm}\Url}\fi
\providecommand{\eprint}[2][]{\url{#2}}

\bibitem[{Al~Mohamad \emph{et~al.}(2022)Al~Mohamad, Goeman, and van
  Zwet}]{al2022simultaneous}
Al~Mohamad D, Goeman JJ, van Zwet EW (2022).
\newblock \enquote{Simultaneous confidence intervals for ranks with application
  to ranking institutions.}
\newblock \emph{Biometrics}, \textbf{78}(1), 238--247.

\bibitem[{Bazylik \emph{et~al.}(2021)Bazylik, Mogstad, Romano, Shaikh, and
  Wilhelm}]{Mogstad:2021bb}
Bazylik S, Mogstad M, Romano JP, Shaikh AM, Wilhelm D (2021).
\newblock \enquote{Finite- and large-sample inference for ranks using
  multinomial data with an application to ranking political parties.}
\newblock \emph{Working Paper CWP40/21}, CeMMAP.

\bibitem[{Boyd and Vandenberghe(2018)}]{boyd2018introduction}
Boyd S, Vandenberghe L (2018).
\newblock \emph{Introduction to applied linear algebra: vectors, matrices, and
  least squares}.
\newblock Cambridge university press.

\bibitem[{Chetty and Hendren(2018)}]{Chetty:2018iu}
Chetty R, Hendren N (2018).
\newblock \enquote{The Impacts of Neighborhoods on Intergenerational Mobility
  I: Childhood Exposure Effects.}
\newblock \emph{The Quarterly Journal of Economics}, \textbf{133}(3),
  1107--1162.

\bibitem[{Chetverikov and Wilhelm(2023)}]{Chetverikov:2023aa}
Chetverikov D, Wilhelm D (2023).
\newblock \enquote{Inference for Rank-Rank Regressions.}
\newblock \emph{Technical report}.

\bibitem[{Deutscher and Mazumder(2023)}]{Deutscher:2023oo}
Deutscher N, Mazumder B (2023).
\newblock \enquote{Measuring Intergenerational Income Mobility: A Synthesis of
  Approaches.}
\newblock \emph{Journal of Economic Literature}, \textbf{61}(3), 988--1036.

\bibitem[{Hofert \emph{et~al.}(2023)Hofert, Kojadinovic, Maechler, and
  Yan}]{copula1}
Hofert M, Kojadinovic I, Maechler M, Yan J (2023).
\newblock \emph{copula: Multivariate Dependence with Copulas}.
\newblock R package version 1.1-2,
  \urlprefix\url{https://CRAN.R-project.org/package=copula}.

\bibitem[{{Ivan Kojadinovic} and {Jun Yan}(2010)}]{copula3}
{Ivan Kojadinovic}, {Jun Yan} (2010).
\newblock \enquote{Modeling Multivariate Distributions with Continuous Margins
  Using the {copula} {R} Package.}
\newblock \emph{Journal of Statistical Software}, \textbf{34}(9), 1--20.
\newblock \urlprefix\url{https://www.jstatsoft.org/v34/i09/}.

\bibitem[{{Jun Yan}(2007)}]{copula2}
{Jun Yan} (2007).
\newblock \enquote{Enjoy the Joy of Copulas: With a Package {copula}.}
\newblock \emph{Journal of Statistical Software}, \textbf{21}(4), 1--21.
\newblock \urlprefix\url{https://www.jstatsoft.org/v21/i04/}.

\bibitem[{Lehmann and Romano(2005)}]{Lehmann:2005p3350}
Lehmann EL, Romano JP (2005).
\newblock \emph{Testing Statistical Hypotheses}.
\newblock Springer, New York.

\bibitem[{{Marius Hofert} and {Martin M\"achler}(2011)}]{copula4}
{Marius Hofert}, {Martin M\"achler} (2011).
\newblock \enquote{Nested Archimedean Copulas Meet {R}: The {nacopula}
  Package.}
\newblock \emph{Journal of Statistical Software}, \textbf{39}(9), 1--20.
\newblock \urlprefix\url{https://www.jstatsoft.org/v39/i09/}.

\bibitem[{Mogstad \emph{et~al.}(2023)Mogstad, Romano, Shaikh, and
  Wilhelm}]{Mogstad:2023aa}
Mogstad M, Romano JP, Shaikh AM, Wilhelm D (2023).
\newblock \enquote{Inference for Ranks with Applications to Mobility across
  Neighbourhoods and Academic Achievement across Countries.}
\newblock \emph{The Review of Economic Studies}, \textbf{forthcoming}.

\bibitem[{Mogstad and Torsvik(2023)}]{Mogstad:2023uu}
Mogstad M, Torsvik G (2023).
\newblock \enquote{Family background, neighborhoods, and intergenerational
  mobility.}
\newblock In S~Lundberg, A~Voena (eds.), \emph{Handbook of the Economics of the
  Family, Volume 1}, volume~1 of \emph{Handbook of the Economics of the
  Family}, chapter~6, pp. 327--387. North-Holland.

\bibitem[{Mohamad \emph{et~al.}(2017{\natexlab{a}})Mohamad, Goeman, and van
  Zwet}]{mohamad2017improvement}
Mohamad DA, Goeman JJ, van Zwet EW (2017{\natexlab{a}}).
\newblock \enquote{An improvement of Tukey's HSD with application to ranking
  institutions.}
\newblock \emph{arXiv preprint arXiv:1708.02428}.

\bibitem[{Mohamad \emph{et~al.}(2017{\natexlab{b}})Mohamad, van Zwet, Goeman,
  and Solari}]{mohamad2017simultaneous}
Mohamad DA, van Zwet EW, Goeman JJ, Solari A (2017{\natexlab{b}}).
\newblock \enquote{Simultaneous confidence sets for ranks using the
  partitioning principle-Technical report.}
\newblock \emph{arXiv preprint arXiv:1708.02729}.

\bibitem[{Nagler and Vatter(2023)}]{rvinecopulib}
Nagler T, Vatter T (2023).
\newblock \emph{rvinecopulib: High Performance Algorithms for Vine Copula
  Modeling}.
\newblock R package version 0.6.3.1.1,
  \urlprefix\url{https://CRAN.R-project.org/package=rvinecopulib}.

\bibitem[{{R Core Team}(2023)}]{RManual}
{R Core Team} (2023).
\newblock \emph{R: A Language and Environment for Statistical Computing}.
\newblock R Foundation for Statistical Computing, Vienna, Austria.
\newblock \urlprefix\url{https://www.R-project.org/}.

\bibitem[{Zeileis(2004)}]{zeileis2004econometric}
Zeileis A (2004).
\newblock \enquote{Econometric computing with HC and HAC covariance matrix
  estimators.}

\bibitem[{Zeileis \emph{et~al.}(2020)Zeileis, K{\"o}ll, and
  Graham}]{zeileis2020various}
Zeileis A, K{\"o}ll S, Graham N (2020).
\newblock \enquote{Various versatile variances: an object-oriented
  implementation of clustered covariances in R.}
\newblock \emph{Journal of Statistical Software}, \textbf{95}, 1--36.

\end{thebibliography}

\newpage

\begin{appendix}

\section{One-Sided Confidence Sets}
\label{app: one-sided}

The main text describes how to construct two-sided marginal and simultaneous confidence sets for the ranks. One-sided confidence sets can be constructed in a similar fashion. For simplicity of exposition, we only show how to construct one-sided simultaneous confidence sets for the ranks with upper endpoints equal to $p$, i.e., they are simultaneous lower confidence bounds on the ranks.

To this end we consider the construction as in \eqref{eq: Rnj} except that the two-sided confidence sets for the differences, $C_{{\rm symm}, n,j,k}$, in the expressions for $N_j^-$ and $N_j^+$ are replaced by the following one-sided confidence sets for the differences:

$$C_{{\rm upper}, n,j,k} := \Biggr (-\infty, \hat{\theta}_j - \hat{\theta}_k + \hat{se}_{jk} c_{{\rm upper},n,j}^{1-\alpha} \Biggr ], $$
where $c_{{\rm upper},n,j}^{1-\alpha}$ is the $(1-\alpha)$-quantile of 
$$\max_{(j,k)\colon k\neq j} \frac{\theta(P_j)-\theta(P_k) - (\hat{\theta}_j - \hat{\theta}_k)}{\hat{se}_{jk}}. $$
As in Section~\ref{sec: simul} the critical value can be approximated by the $(1-\alpha)$-quantile of the $m$ draws of $\max_{(j,k)\colon k\neq j} (Z_k-Z_j)/\hat{se}_{jk}$.

\section{Efficient Computation}
\label{app:technical}

This appendix provides more details on the implementation of the inference methods for rank-rank regressions by \cite{Chetverikov:2023aa}, in particular on the efficient computation of standard errors. Consider the OLS estimator $\hat{\rho}$ in \eqref{eq: joint ols estimator}. Its asymptotic variance can be written as
\begin{equation}\label{eq: our general variance formula}
    \Sigma_{1,1}  := \frac{1}{\sigma_{\nu}^4} E\Big[(h_1(X,W,Y) + h_2(X,Y) + h_3(X))^2\Big] 
    \end{equation}
    with $\sigma_{\nu}^2 := E[\nu^2]$, and
    \begin{align*}
        I(u,v) &:= \omega 1\{u\leq v\} + (1-\omega)1\{u<v\},\\
        h_1(x,w,y) &:= (R_Y(y) - \rho R_X(x) - w'\beta)(R_X(x) - w'\gamma),\\
        h_2(x,y) & := E[(I(y,Y) - \rho I(x,X) - W'\beta)(R_X(X) - W'\gamma)],\\
        h_3(x) & := E[(R_Y(Y) - \rho R_X(X) - W'\beta)(I(x,X) - W'\gamma)].
    \end{align*}
A consistent estimator of this asymptotic variance (\citet[Lemma 4]{Chetverikov:2023aa}) is
\begin{equation}
\label{eqn:B-var-est}
\hat{\Sigma}_{1,1} := \frac{1}{n \hat{\sigma}_{\nu}^4} \sum_{i=1}^n(H_{1i} + H_{2i} + H_{3i})^2, 
\end{equation}
where $\hat{\sigma}_{\nu}^2 := n^{-1}\sum_{i=1}^n \hat{\nu}_i^2$ is an empirical analog of $\sigma_{\nu}^2 = E[\nu^2]$, $\hat{\nu}_i := R_i^X-W_i'\hat{\gamma}$, and
\begin{align*}
    H_{1i} &:= \left(R_i^Y - \hat{\rho} R_i^X - W_i'\hat{\beta}\right)\left(R_i^X - W_i'\hat{\gamma}\right),\\
    H_{2i} & := \frac{1}{n}\sum_{j=1}^n \left(I(Y_i,Y_j) - \hat{\rho} I(X_i,X_j)-W_j'\hat{\beta}\right)\left(R_j^X-W_j'\hat{\gamma}\right),\\
    H_{3i} & := \frac{1}{n}\sum_{j=1}^n \left(R_j^Y - \hat{\rho} R_j^X - W_j'\hat{\beta}\right)\left(I(X_i,X_j) - W_j'\hat{\gamma}\right).
\end{align*}
The usual workflow in \proglang{R} is to first run the main rank-rank regression (i.e. estimate $\hat{\rho}$ and $\hat{\beta}$) and then estimate the variance, e.g. by calling \code{summary}. Thus it is assumed, that the estimates $\hat{\rho}$ and $\hat{\beta}$ are available before computing the estimator \eqref{eqn:B-var-est}. Then the computation of the estimator proceeds in five steps:
\begin{enumerate}
    \item Computation of the projection coefficients $\hat{\gamma}$ from an OLS regression of $R_i^X$ on $W_i$.
    \item Computation of the residuals from regression of $R_i^X$ on $W_i$.
    \item Computation of the residual variance $\hat{\sigma}_{\nu}^2$.
    \item Computation of $H_{1i}$, $H_{2i}$, and $H_{3i}$.
    \item Substitution into the formula for $\hat{\Sigma}_{1,1}$.
\end{enumerate}
The computation of the other diagonal elements $\hat{\Sigma}_{j,j}$, $j>1$, the estimators of the asymptotic variances of the other regression coefficients in the vector $\hat{\rho}$, is analagous and, thus, potentially requires repeatedly executing computations similar to Steps 1. -- 5. for each $j=1,\ldots,p+1$, where $p$ is the dimension of $W_i$. In the following two subsections, we describe how the implementation in \pkg{lmranks} avoids unnecessary repeated calculation from scratch of the projection coefficients (Step 1.) and simplifies the computations of $H_{1i}$, $H_{2i}$, and $H_{3i}$ (Step 4.). 

Let us analyze the computational complexity of the task of computing the asymptotic variances for all regression coefficients. As shown in the following subsections, the complexity of Step 1. is $O(p^3)$ and that of Step 4. is $O(n log(n) + np)$.
The complexity of Step 2. is $O(np)$ for one coefficient and $O(np^2)$ for all coefficients. The complexity of Step 3. is $O(n)$ for one coefficient and $O(np)$ for all coefficients. Similarly, the complexity of Step 5. is $O(n)$ for one coefficient and $O(np)$ for all coefficients. To summarize, the complexity of the entire procedure is $O(p^3 + np^2 + np + (nlog(n) + np) + np)=O(p^3+np^2+nlog(n))$. Thus it is linearithmic with respect to the sample size $n$.

\subsection{Efficient Computation of the Projection Coefficients in Step 1.}

Denote by $Z_i := (R_i^X,W_i')'$ the $(p+1)$-dimensional vector collecting all regressors and by $Z$ the $(n\times (p+1))$ matrix with $i$-th row equal to $Z_i$. 
Denote by $Z_{(j)}$ the $j$-th column of $Z$, by $Z_{(-j)}$ the matrix $Z$ after removing the $j$-th column. Let $\hat{\gamma}_{j,k}$ be the $k$-th coefficient from a regression of $Z_{(j)}$ on to $Z_{(-j)}$, and let $\hat{\gamma}_j := (\hat{\gamma}_{j,1},\ldots,\hat{\gamma}_{j,p+1})'$. Then,
\begin{equation}\label{eq: proj coefs}
    \hat{\gamma}_j = \left(Z_{(-j)}'Z_{(-j)}\right)^{-1} Z_{(-j)}'Z_{(j)},
\end{equation}
which is a well-known result in OLS regression.
We now show that it is not necessary to compute this expression separately for each $j=1,\ldots,p+1$. Partition $Z$ into two blocks, $Z =[ Z_{(-(p+1))}\; Z_{(p+1)}]$. Then, using the block matrix inverse of $Z'Z$, one can show that 
\begin{equation*}
 (Z'Z)^{-1}_{(p+1)} = \begin{pmatrix}
 -\hat{\gamma_{p+1}}'c \\
 c
 \end{pmatrix},
\end{equation*}
where the scalar $c$ is defined as $c:=(Z'Z)^{-1}_{p+1,p+1}$.
Then, letting $D:=diag((Z'Z)^{-1})$ be a diagonal matrix containing the diagonal of $(Z'Z)^{-1}$, one can show that
\begin{equation}\label{eq: insight}
    (Z'Z)^{-1} D^{-1} = \begin{pmatrix}
  1 & -\hat{\gamma}_{2,1} & \cdots & -\hat{\gamma}_{p+1,1} \\
  -\hat{\gamma}_{1,1} & 1 & \cdots & -\hat{\gamma}_{p+1,2} \\
  \vdots  &  \vdots & \ddots& \vdots \\
  -\hat{\gamma}_{1,p} & -\hat{\gamma}_{2, p} & \cdots & 1
\end{pmatrix}.
\end{equation}
So, we can calculate all projection coefficients using one operation, namely the inversion of $Z'Z$. For this purpose we use the QR factorization of $Z$, which has already been calculated and stored in memory from the estimation of the rank-rank regression. The factorization yields the Cholesky decomposition of $Z'Z$, which can be passed into the function \code{chol2inv} from base \proglang{R} for the calculation of the inverse with computational complexity $O(p^3)$ (\cite{boyd2018introduction}).

A naive implementation would require the calculation of the expression \ref{eq: proj coefs} (or, equivalently, fitting each projection model) separately for each $j=1,\ldots,p+1$. In \proglang{R}, fitting a linear model is achieved by finding a QR decomposition of the matrix with explanatory variables and back substitution (\cite{RManual}). The first step, computationally more complex than the second one, is $O(np^2)$ (\cite{boyd2018introduction}). Thus, a naive implementation would require finding QR factorizations of the matrices $Z_{(-j)}$ for each $j=1,\ldots,p+1$, which would be $O(np^2(p+1))=O(np^3)$.

In contrast, the insight in \eqref{eq: insight} allows us to compute the projection coefficients using the QR decomposition of $Z$ already obtained in the main rank-rank regression. The only new step is finding the inverse of $Z'Z$ (and other minor calculations) which requires $O(p^3)$ operations. Thus, the complexity of the new procedure is constant with respect to sample size $n$.

\subsection{Efficient Computation of the Summands in Step 4.}

The computation of the estimator $\hat{\Sigma}_{1,1}$ of the asymptotic variance $\Sigma_{1,1}$ requires calculation of the terms $H_{1i}$, $H_{2i}$, and $H_{3i}$ for each $i=1,\ldots,n$. Each of the terms $H_{2i}$ and $H_{3i}$ includes a sum iterating over $j=1,\ldots,n$. The number of operations needed for a naive approach using nested for-loops would scale as $O(n^2)$ and thus lead to time-consuming calculations when the sample size is large. The goal of this subsection is to explain how the three terms can be computed with fewer than $O(n^2)$ operations.

We illustrate the approach by focusing on the term $H_{3i}$, which can be written as 
$$H_{3i} = \underbrace{\frac{1}{n}\sum_{j=1}^n \left(R_j^Y - \hat{\rho} R_j^X - W_j'\hat{\beta}\right)I(X_i,X_j)}_{=: \tilde{H}_{3i}} - \frac{1}{n}\sum_{j=1}^n \left(R_j^Y - \hat{\rho} R_j^X - W_j'\hat{\beta}\right) W_j'\hat{\gamma}.$$
The second term depends only on $j$ and thus the number of operations required to compute it is linear with respect to sample size $n$. The first term, $\tilde{H}_{3i}$, is a sum over $j$ that has to be computed for each $i$ and is thus computationally more challenging. 

Let $\hat{\varepsilon} := (\hat{\varepsilon}_1,\ldots,\hat{\varepsilon}_n)'$ with $\hat{\varepsilon}_i := R_i^Y - \hat{\rho} R_i^X - W_i'\hat{\beta}$ be the vector of residuals from the rank-rank regression, and denote by $\mathbf{I}$ the matrix with $(i,j)$ element equal to $I(X_i,X_j)$. Then, 
$$ \begin{pmatrix} \tilde{H}_{31}\\ \vdots\\ \tilde{H}_{3n} \end{pmatrix} = \mathbf{I} \hat{\varepsilon}. $$
To see how this matrix product can be computed with fewer than $O(n^2)$ operations, first, consider the special case in which the observations are ordered such that $X_1 > X_2 > \ldots > X_n$. Then, $\mathbf I$ is a lower-triangular matrix with ones below the diagonal, $\omega$ on the diagonal, and zeroes above the diagonal.
Define the \textit{cumulative sum} of vector $v=(v_1,\ldots,v_n)'$ as $(v_1, v_1+v_2, v_1+v_2+v_3,\ldots,\sum_{i=1}^nv_i)'$.  Then $\mathbf{I} \hat{\varepsilon}$ is equivalent to the cumulative sum of the vector
$$ \omega \hat{\varepsilon} + (1-\omega) \begin{pmatrix} 0\\ \hat{\varepsilon}_1 \\ \vdots \\ \hat{\varepsilon}_{n-1} \end{pmatrix}. $$
Cumulative sums are implemented in base \proglang{R} as \fct{cumsum} and require $O(n)$ operations, which is faster than the naive implementation using a loop to compute $n$ sums of $n$ terms each.

Now, consider the case in which there may be ties, i.e. $X_1 \geq X_2 \geq \ldots \geq X_n$. Then, the above representation of $\mathbf{I} \hat{\varepsilon}$ as cumulative sum does no longer hold. However, a similar approach still applies after noting that $\mathbf{I}$ is a weighted average of two indicators, each defining a matrix of indicators with a cumulative sum representation. The vector $\hat\varepsilon$ only needs to be suitably prepared before applying the cumulative sum. 
Briefly speaking, this is done by grouping the entries in $\hat\varepsilon$ corresponding to duplicates in $X$. For each group of such entries, a sum is calculated, then the first (for $\omega=1$) or last (for $\omega = 0$) element is replaced with the calculated sum and the rest with 0.

Finally, the general case when $X_1,\ldots,X_n$ are not ordered can simply be reduced to the case $X_1 \geq X_2 \geq \ldots \geq X_n$ by first sorting, then applying the previous approach to the sorted observations, and finally reversing the sorting. 

The whole procedure is summarised in Algorithm \ref{alg:ind-mat-mult}. The operations from lines 1 and 24 are $O(nlog(n))$ each. The operations from lines 2, 10, 11, 21, 23 and 25 as well as from whole blocks 4-9 and 12-21 are $O(n)$ each. Therefore, the algorithmic complexity of Algorithm \ref{alg:ind-mat-mult} is $O(nlog(n))$. This is a favorable complexity in comparison to $O(n^2)$ of the naive approach, especially for large $n$.

It is worth noting that this approach is easily extendable for the task of computing all the estimators $\hat\Sigma_{1,j}$ for $j=1,...,p+1$. Then, the sorting permutation of $X$ needs to be found only once (and thus the operations from lines 1 and 24 of Algorithm \ref{alg:ind-mat-mult} are computed only once). The complexities of the remaining operations of Algorithm \ref{alg:ind-mat-mult} are linear with respect to sample size $n$. Thus, the complexity of computing the terms $H_{1,i}, H_{2,i}$ and $H_{3,i}$ for all $p+1$ asymptotic variance estimators is $O(nlog(n) + np)$.

\begin{algorithm}
\caption{Multiplication of indicator matrix of vector $x$ with arbitrary vector $v$}
\label{alg:ind-mat-mult}
\begin{algorithmic}[1]
\Require $n,p\in\mathbb N, \omega \in [0,1], x\in\mathbb R^n, v \in\mathbb R^{n}$.
\State Find a sorting permutation $\sigma$ of vector $x$.
\State Apply the $\sigma$ to $x$ and $v$, in place.
\For{$\tilde \omega = 1,0$}
\If{$\tilde \omega = 0$}
    \Comment{Shift $v$ by one place because of 0s on diagonal of $\mathbf I$}.
    \For{$i = n,n-1,...,2$}
        \State $v_{i} \gets v_{i-1}$
    \EndFor
    \State $v_{1} \gets 0$
\EndIf
\State Identify the set of groups $G$ of elements of $v$ corresponding to equal entries in $x$.
\State $\tilde v \gets v$
\For{$g$ in $G$}
    \State Calculate sum $\mathbf s$ of $v$'s elements in group $g$.
    \State Set $0$ in all $\tilde v$'s elements in group $g$.
    \If{$\tilde \omega = 0$}
        \State Set the $\tilde v$'s first element in group $g$ to $\mathbf s$.
    \Else 
        \State Set the $\tilde v$'s last element in group $g$ to $\mathbf s$.
    \EndIf
\EndFor
\State $c_{\tilde \omega} \gets$ cumulative sum of $\tilde v$
\EndFor
\State $c \gets \omega c_1 + (1-\omega)c_0$ 
\Comment Multiplication and addition element-wise.
\State Find the inverse of $\sigma$.
\State Apply the $\sigma^{-1}$ to $c$.
\State \Return $c$
\end{algorithmic}
\end{algorithm}

\end{appendix}

\end{document}